\documentclass[aps,prl,reprint,superscriptaddress,floatfix]{revtex4-2} 
\usepackage{amsmath}
\usepackage{graphicx}
\usepackage{hyperref}
\usepackage[T1]{fontenc} 
\usepackage{svg}

\begin{document}

\title{Photoinduced Frustration Modulation in $\kappa$-type Quantum Spin Liquid Candidates}

\author{M. Tepie}
\thanks{These authors contributed equally to this work}
\author{F. Glerean}
\thanks{These authors contributed equally to this work}
\affiliation{Department of Physics, Harvard University, Cambridge, Massachusetts 02138, USA\looseness=-1}

\author{J. Ov\v{c}ar}
\affiliation{Division of Theoretical Physics, Ruđer Bošković Institute, Zagreb 10000, Croatia}

\author{S. Priya}
\affiliation{1. Physikalisches Institut, Universität Stuttgart, Stuttgart 70569, Germany}

\author{K. Miyagawa}
\affiliation{Department of Applied Physics, University of Tokyo, Tokyo 113-8656, Japan}

\author{H. Taniguchi}
\affiliation{Graduate School of Science and Engineering, Saitama University, Saitama 338-8570, Japan}

\author{K. Kanoda}
\affiliation{1. Physikalisches Institut, Universität Stuttgart, Stuttgart 70569, Germany}
\affiliation{Department of Advanced Materials Science, University of Tokyo, Kashiwa 277-8561, Japan}
\affiliation{Max Planck Institute for Solid State Research, 70569 Stuttgart, Germany}

\author{I. Lon\v{c}ari\'{c}}
\affiliation{Division of Theoretical Physics, Ruđer Bošković Institute, Zagreb 10000, Croatia}

\author{M. Dressel}
\affiliation{1. Physikalisches Institut, Universität Stuttgart, Stuttgart 70569, Germany}

\author{M. Mitrano}
\email{mmitrano@fas.harvard.edu}
\affiliation{Department of Physics, Harvard University, Cambridge, Massachusetts 02138, USA\looseness=-1}

\date{\today}

\begin{abstract}  

Geometric frustration is a key parameter controlling electronic and magnetic properties of quantum spin liquid systems, yet remains challenging to tune. Here, we coherently drive molecular vibrations with midinfrared pulses in two organic quantum spin liquid candidates, the insulating $\kappa$-(BEDT-TTF)$_2$Cu$_2$(CN)$_3$ and the metallic $\kappa$-(BEDT-TTF)$_4$Hg$_{2.89}$Br$_8$, and probe their electronic response through ultrafast reflectivity measurements. We observe a nonlinear coupling between local molecular vibrations and nonlocal phonons, which is expected to directly modulate the geometric frustration of their triangular lattice. Our findings establish a promising route to dynamically control frustration in nonbipartite quantum materials.

\end{abstract}

\maketitle 

\begin{figure*}[t]
    \centering
    \includegraphics[width=.75\textwidth]{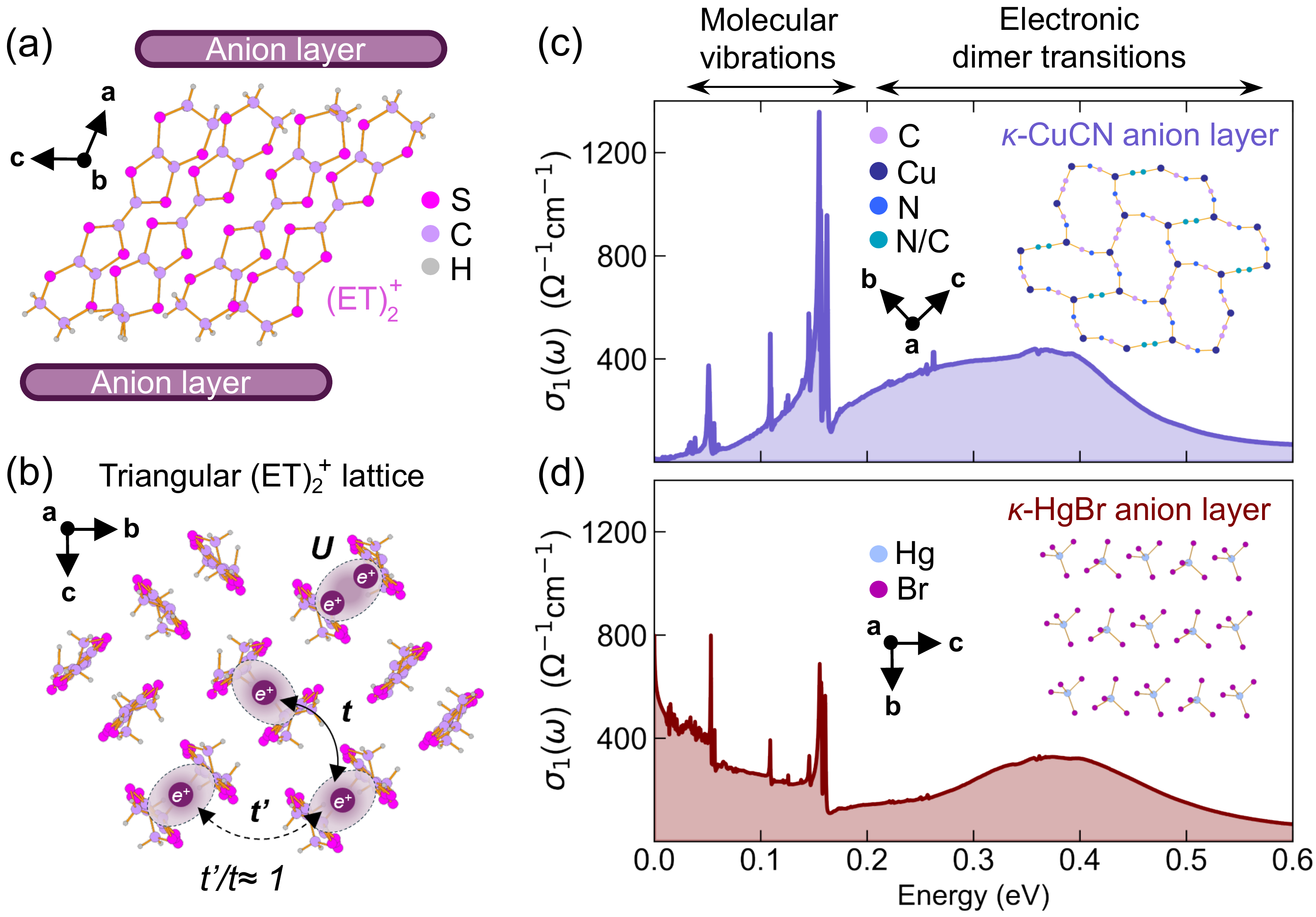}
    \caption{(a) Alternating cation and anion layers stacked along the \emph{a} axis in the $\kappa$-type structure of ET-based organic charge-transfer salts. (b) In the cation layer, (ET)$_2^+$ dimers with effective charge $+e$ and spin $S = 1/2$ are arranged in a triangular lattice. (c)-(d) Equilibrium optical conductivity spectra of $\kappa$-CuCN and $\kappa$-HgBr measured at 100 K along the {\it b}-axis. The optical response depends critically on the structure of the anion layers. At $\omega\rightarrow 0$, $\kappa$-CuCN exhibits insulating behavior, whereas $\kappa$-HgBr manifests a Drude peak. At higher energy, the response is characterized by sharp molecular modes and charge-transfer excitations of the molecular dimers.}
    \label{fig:fig1}
\end{figure*}

Organic charge-transfer salts $\kappa$-(BEDT-TTF)$_2$X are prototypical correlated electron systems exhibiting Mott insulating, superconducting, and possible quantum spin liquid (QSL) states \cite{Ishiguro1998organic,Dressel2004optical,Seo2004toward,Kanoda2011Mott,Powell2011quantum,Menke2024superconductivity}. These quasi-two-dimensional compounds consist of alternating layers of BEDT-TTF (bisethylenedithio-tetrathiafulvalene, abbreviated ET) molecules and anionic charge reservoirs [Fig.~\ref{fig:fig1}(a)]. Within the ET layers, strong dimerization and charge transfer to the anions yield one hole per dimer, effectively realizing a half-filled conduction band with strong Coulomb repulsion \cite{Ishiguro1998organic}. As a result, they are well described by a Hubbard model on an anisotropic triangular lattice [Fig.~\ref{fig:fig1}(b)] with onsite Coulomb repulsion $U$, and inequivalent hopping amplitudes $t$ and $t^{\prime}$ \cite{Powell2011quantum,Koretsune2014evaluating,Kandpal2009revision}. The triangular geometry, with its inherent frustration and low coordination number, makes them an ideal platform to probe strong quantum fluctuations and emergent QSL behavior.

The insulating QSL candidate $\kappa$-(BEDT-TTF)$_2$Cu$_2$(CN)$_3$ ($\kappa$-CuCN) is among the most compelling members of this material class. Its $S=1/2$ spins reside on a highly frustrated dimer lattice and interact via an antiferromagnetic exchange $J = 250$ K, yet remain magnetically disordered down to 32 mK \cite{Shimizu2003spin}. This behavior has been interpreted as a gapless QSL with deconfined spinon excitations \cite{Balents2010spin}, although recent electron spin resonance experiments point to a spin gap, possibly reflecting a valence bond solid stabilized by magnetoelastic coupling \cite{Miksch2021gapped}. A complete anion substitution, resulting in the incommensurate structure $\kappa$-(BEDT-TTF)$_4$Hg$_{2.89}$Br$_8$ ($\kappa$-HgBr)\cite{Lyubovskaya1987organic,Lyubovskaya1987superconductivity,Lyubovskaya1991controlled}, drives the system into a metallic state with non-Fermi liquid–like transport behavior and a superconducting phase below 4–6 K \cite{Taniguchi2007anomalous,Oike2015,Oike2017,Suzuki2022}. With comparable lattice geometry but distinct ground states, $\kappa$-CuCN and $\kappa$-HgBr offer a unique opportunity to investigate how superconductivity and magnetic order emerge from a magnetically disordered parent state, providing an alternative to the antiferromagnetic route observed in high-$T_c$ cuprates.

Electron correlations ($U/t$) and geometric frustration ($t^{\prime}/t$) are the key parameters governing the ground states of $\kappa$-type materials. Beyond equilibrium control via chemical substitution and pressure \cite{Pustogow2021rise, Pustogow2023chasing}, ultrafast optical methods offer a powerful means to dynamically manipulate electronic correlations. In particular, nonresonant excitation of intradimer molecular displacements has been shown to induce an insulator-to-metal transition by modulating the dimer Coulomb repulsion \cite{Kawakami2009optical}. Strong electron-molecular vibration coupling in these systems similarly enables control over the Hubbard $U$ through selective excitation of vibrational modes. Following light control experiments in one-dimensional Mott insulators \cite{Kaiser2014optical,Singla2015THz}, this approach has recently led to the observation of a nonequilibrium superconducting response in metallic $\kappa$-(ET)$_2$Cu[N(CN)$_2$]Br \cite{buzzi2020photomolecular}. Whether such excitation can also modulate geometric frustration, a key parameter governing magnetic interactions, remains unexplored.

In this Letter, we demonstrate a route to ultrafast control of geometric frustration in $\kappa$-(ET)$_2$X compounds via selective excitation of molecular vibrations. We use intense midinfrared pulses, resonantly tuned to local ET vibrational modes, to drive both insulating $\kappa$-CuCN and metallic $\kappa$-HgBr, while monitoring the dimer response via transient reflectivity measurements. We find that the excited local modes couple to $\sim$1.2 THz Raman-active phonons involving dimer breathing and rotation, accompanied by cooperative motion of the anionic and cationic layers. These coherent lattice dynamics are predicted to induce substantial changes in both the correlation strength $U/t$ and, critically, the effective frustration of the triangular lattice.

Fig.~\ref{fig:fig1}(c),(d) shows the equilibrium optical spectra of $\kappa$-CuCN and $\kappa$-HgBr single crystals, measured using Fourier-transform infrared spectroscopy. The two compounds differ primarily in their anion layers and exhibit distinct optical conductivities in the \emph{bc}-plane. Both systems share high-energy absorption features around 0.3–0.4 eV, corresponding to interdimer and intradimer charge-transfer transitions \cite{Dressel2004optical}. At mid-infrared frequencies, the spectra display multiple sharp peaks arising from infrared-active vibrational modes of the ET molecules \cite{Dressel2016lattice}. However, $\kappa$-CuCN is insulating, with vanishing in-gap optical conductivity for $\omega\rightarrow0$, while metallic $\kappa$-HgBr exhibits pronounced Drude-like carrier excitations \cite{supp}. Notably, the in-gap absorption of the QSL candidate $\kappa$-CuCN exceeds expectations for a conventional Mott insulator \cite{Kezsmarki2006depressed,Elsasser2012power}. Its origin has been variously attributed to gapless spinons, a paired-electron liquid, dipolar fluctuations, or quantum criticality \cite{Ng2007power,Hotta2010quantum,Pustogow2022thirty,Liebman2024novel}.

\begin{figure}[t]
    \centering
    \includegraphics[width=.47\textwidth]{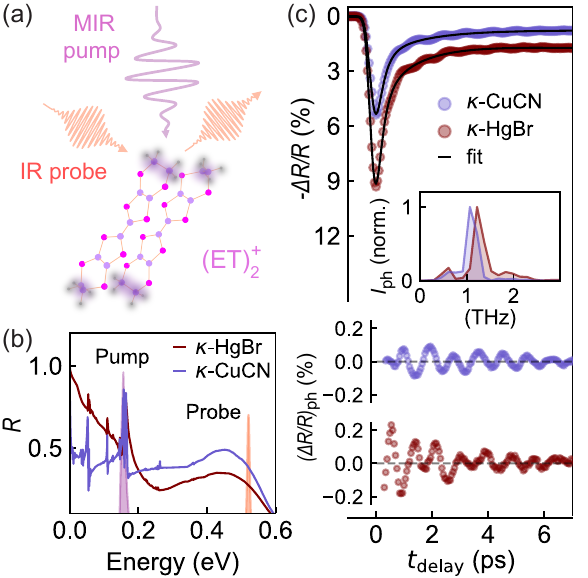}
    \caption{(a) Sketch of the midinfrared (MIR) pump - near-infrared (NIR) probe experiment targeting the (ET)$_2^+$ dimers. (b) Pump and probe spectra (shaded areas) centered on the ethylene group vibrations and the intradimer absorption band of $\kappa$-CuCN and $\kappa$-HgBr, respectively. Solid curves are the sample reflectivities ({\it b}-axis, 100 K) (c) Time-dependent reflectivity change of the intradimer resonance (symbols). Both compounds exhibit a transient electronic response (top) following a double exponential decay (black solid lines) together with a coherent oscillatory behavior, which we isolate by subtracting the exponential contribution (bottom). The oscillations are approximately centered at 1.2 THz (inset, Fourier-transform spectrum).}
    \label{fig2}
\end{figure}

We excite $\kappa$-CuCN and $\kappa$-HgBr samples with femtosecond midinfrared pump pulses and probe their optical response at near-infrared frequencies [Fig.~\ref{fig2}(a)]. We generate carrier-envelope-phase stable pump pulses via difference frequency generation in a 600-$\mu$m GaSe crystal, using signal beams from two optical parametric amplifiers (OPAs) pumped by an amplified Ti:sapphire laser (800 nm, 1 kHz) and seeded by a common white-light continuum. The resulting 200-fs mid-infrared pulses are polarized along the in-plane crystallographic axes and tuned at resonance with ethylene end-group modes [Fig.~\ref{fig2}(b)], which exhibit pronounced Fano profiles, indicative of strong electron-molecular vibration coupling. We probe the intradimer response by monitoring reflectivity changes at 520 meV along the {\it b}-axis using the idler output of one OPA and a fast photodiode. All measurements are conducted at 100 K in a cryostat, deep in the insulating and metallic phases of $\kappa$-CuCN and $\kappa$-HgBr, respectively.

Figure~\ref{fig2}(c) displays time-resolved reflectivity changes at the intradimer absorption following excitation with 161 meV pump photons at a fluence of 1~mJ/cm$^2$. We observe a prompt reflectivity decrease of 5\% in $\kappa$-CuCN and 10\% in $\kappa$-HgBr, followed by a rapid double-exponential recovery. This transient reduction reflects a loss of intradimer charge-transfer spectral weight, consistent with a light-induced change of electronic parameters \cite{Singla2015THz,Kawakami2009optical,Tindall2020dynamical}, the impulsive creation of doublon–holon pairs \cite{Oka2005ground,Oka2012nonlinear,Yamakawa2017Mott, Yamakawa2021}, or a combination of the two mechanisms. Recovery occurs on two distinct timescales. We attribute the fast component $\tau_1 \simeq 0.2$ ps to electron–molecular vibration energy transfer and the slower component $\tau_2 \simeq 1.2$–1.5 ps to thermalization of high-energy vibrational modes \cite{Mitrano2014pressure,Singla2015THz,buzzi2020photomolecular}. The amplitude of this reflectivity change scales linearly with fluence up to $\sim$3 mJ/cm$^2$ and increases gradually upon cooling (see Supplemental Material \cite{supp}).

The electronic response of the dimers is accompanied by a coherent oscillation, which constitutes the main observation of this work. By subtracting the electronic background, we isolate and Fourier-transform the phonon reflectivity change $(\Delta R/R)_{ph}$ [Fig.~\ref{fig2}(c)], yielding a peak located at 1.12 THz in $\kappa$-CuCN and 1.18 THz in $\kappa$-HgBr. We characterize the coherent mode by analyzing its dependence on pump energy and probe polarization. We define an effective photosusceptibility $\chi_{ph} = |d\rho/df|$, where $\rho = (\Delta R/R)_{ph}$ and $f = F(1 - R)$ is the absorbed pump fluence ($F$ being the incident pump fluence) \cite{Caviglia2012ultrafast}. As reflectivity changes scale linearly with fluence (see Supplementary \cite{supp}), we probe $\chi_{ph}$ as a function of pump energy and find it strongly enhanced at resonance with vibrations of the ethylene group in the absorption spectrum $\alpha$ of both $\kappa$-CuCN and $\kappa$-HgBr [Fig.~\ref{fig3}(a)-(b)]. We note that the electronic response is similarly resonant with these vibrations (see Supplementary information \cite{supp}, Section IIB). The probe polarization dependence shows an isotropic photosusceptibility in $\kappa$-CuCN and a $C_2$-symmetric pattern in $\kappa$-HgBr [Fig.~\ref{fig3}(c)-(d)]. While both compounds share identical layered cation structures, their anion arrangements differ: $\kappa$-CuCN features an isotropic anion network, whereas $\kappa$-HgBr exhibits a directional, $C_2$-symmetric stripe-like pattern. Therefore, the $\sim$1.2 THz mode modulating the intradimer optical response must involve a cooperative motion of both cation and anion layers, in which the anisotropy is transferred to the BEDT-TTF layers by the coupling via ethylene endgroups and hydrogen bonds \cite{Alemany2015structural,Pouget2018donor}. This coherent response is found to persist upon cooling down to the lowest temperatures accessed in our measurements (see Supplementary Information \cite{supp} Section IIC).

\begin{figure}[t]
    \centering
    \includegraphics[width=.45\textwidth]{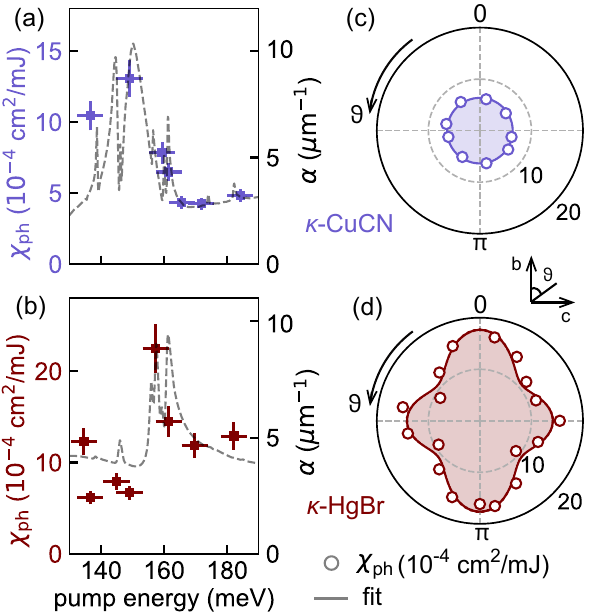}
    \caption{Effective phonon photosusceptibility (symbols) and linear absorption $\alpha$ (dashed lines) as a function of pump photon energy for $\kappa$-CuCN (a) and $\kappa$-HgBr (b). The oscillatory response resonates with the ethylene molecular modes. The pump is polarized along the $c$- and $b$-axes, respectively. The probe polarization dependence of the effective phonon susceptibility (measured with 161 meV pump along $b$) is isotropic for $\kappa$-CuCN (c) and anisotropic for $\kappa$-HgBr (d), reflecting the distinct anion lattice symmetries.}
    \label{fig3}
\end{figure}

Our observations are consistent with the nonlinear coupling between a \emph{local} molecular vibration and a \emph{nonlocal} phonon mode \cite{Forst2011nonlinear,Disa2021engineering}. The coherent mode is an $A_1$-symmetry phonon with mixed infrared and Raman character, involving both cation and anion layers \cite{Dressel2016lattice,Liebman2024novel}. It is strongly anharmonic and coupled to electronic states, as evident from its pronounced broadening and Fano profile in equilibrium Raman \cite{Liebman2024novel} and infrared spectra \cite{Dressel2016lattice}. Its resonant photosusceptibility enhancement underscores a direct coupling with the driven ethylene vibrations. The lowest allowed nonlinear phonon coupling is cubic, of the form $Q_{\mathrm{IR}}^2 Q_{\mathrm{R}}$\cite{Forst2011nonlinear,Disa2021engineering}, where $Q_{\mathrm{IR}}$ is the driven infrared-active molecular mode and $Q_{\mathrm{R}}$ is the coupled $A_1$ phonon mode. Within a $C_{2v}$ symmetry description, each CH$_2$ ethylene mode ($A_1$, $B_1$, or $B_2$) \cite{Kozlov1987assignment} can quadratically couple to the $\sim$1.2 THz $A_1$ phonon, in agreement with the $|E_{\mathrm{pump}}|
^2$ dependence of the $A_1$ mode amplitude \cite{supp}. Intuitively, this interaction is mediated by the ethylene end groups, which act as structural hinges, extending distortions of the BEDT-TTF molecules into the anion layers. Beyond the coherent oscillations, this nonlinear coupling also entails a quasistatic, displacive structural response corresponding to a rectification of the mode coordinate \cite{Forst2011nonlinear,Subedi2014theory}, which should contribute to the initial exponential response, consistent with the energy-dependent photosusceptibility in Supplementary section IIB \cite{supp}. Finally, we note the presence of a finite susceptibility persisting when detuning the pump energy away from the ethylene modes. This background $\chi_{ph}$ suggests the presence of a stimulated Raman scattering contribution in addition to the $Q_{\mathrm{IR}}^2 Q_{\mathrm{R}}$ coupling, in agreement with prior off-resonant optical experiments \cite{Itoh2013collective}.

To determine the electronic parameters of the transiently distorted structure, we perform {\it ab initio} band structure calculations within the frozen-phonon approximation (see Supplementary \cite{supp}). Density functional theory (DFT) calculations reveal multiple phonons in the 1.0–1.3 THz range, involving coupled displacements of anion layers and ET dimers in both compounds. Among these, modes featuring an in-phase modulation of the dimers impart the largest variation in the Hubbard parameters (see Supplementary \cite{supp}). Figures \ref{fig4}(a)-(b) illustrate the phonon-displaced ET layers: in $\kappa$-CuCN, the motion resembles a breathing of the dimers, while in $\kappa$-HgBr it also includes an azimuthal rotation around their axis. The rotation is uniquely enabled by the relative out-of-plane displacement of the two anionic sublattices. Intuitively, a larger dimer size reduces the onsite Coulomb repulsion $U$, while increased proximity between dimers enhances the hopping integrals $t$ and $t'$. To extract phonon-induced changes in electronic correlations and geometric frustration, we fit the displaced band structures to a tight-binding Hubbard model. Note that, to be able to investigate the dependence of the Hubbard parameters on displacements along phonon coordinates, we calculate the parameters around fully relaxed local DFT structural minima, while previous investigations commonly employed experimentally obtained or partially relaxed structures \cite{Kandpal2009revision, nakamura2009, jeschke2010, jeschke2012, Koretsune2014evaluating, guterding2015, Guterding2016} (see Supplementary \cite{supp} for a detailed discussion on the dependence of the calculated hopping parameters on the crystal structure). Despite the mixed character of the nonlinearly excited phonon, which involves both cation and anion motion, the low-energy bands retain dominant ET character, justifying a triangular lattice description with renormalized Hubbard parameters, consistent with prior \emph{ab initio} calculations \cite{jeschke2012} and with the universal $\kappa$-(ET)$_2$X phase diagram upon anion substitution at equilibrium \cite{Powell2011quantum,Dressel2004optical}. As shown in Fig.\ref{fig4}(c), the frustration ratio $t'/t$ varies linearly with the displacement along the $\sim$1.2 THz phonon coordinate \cite{supp}. Interestingly, while both the effective correlation strength $U/t$ and $t'/t$ scale proportionally in $\kappa$-CuCN, $U/t$ is only slightly affected by the phonon displacement in $\kappa$-HgBr. We note that the ratio between intradimer and interdimer distance is strongly modified by the phonon displacement in $\kappa$-CuCN, while it is minimally perturbed in $\kappa$-HgBr, where dimer rotations dominate the mode dynamics. 

\begin{figure}[h]
    \centering
    \includegraphics[width=.45\textwidth]{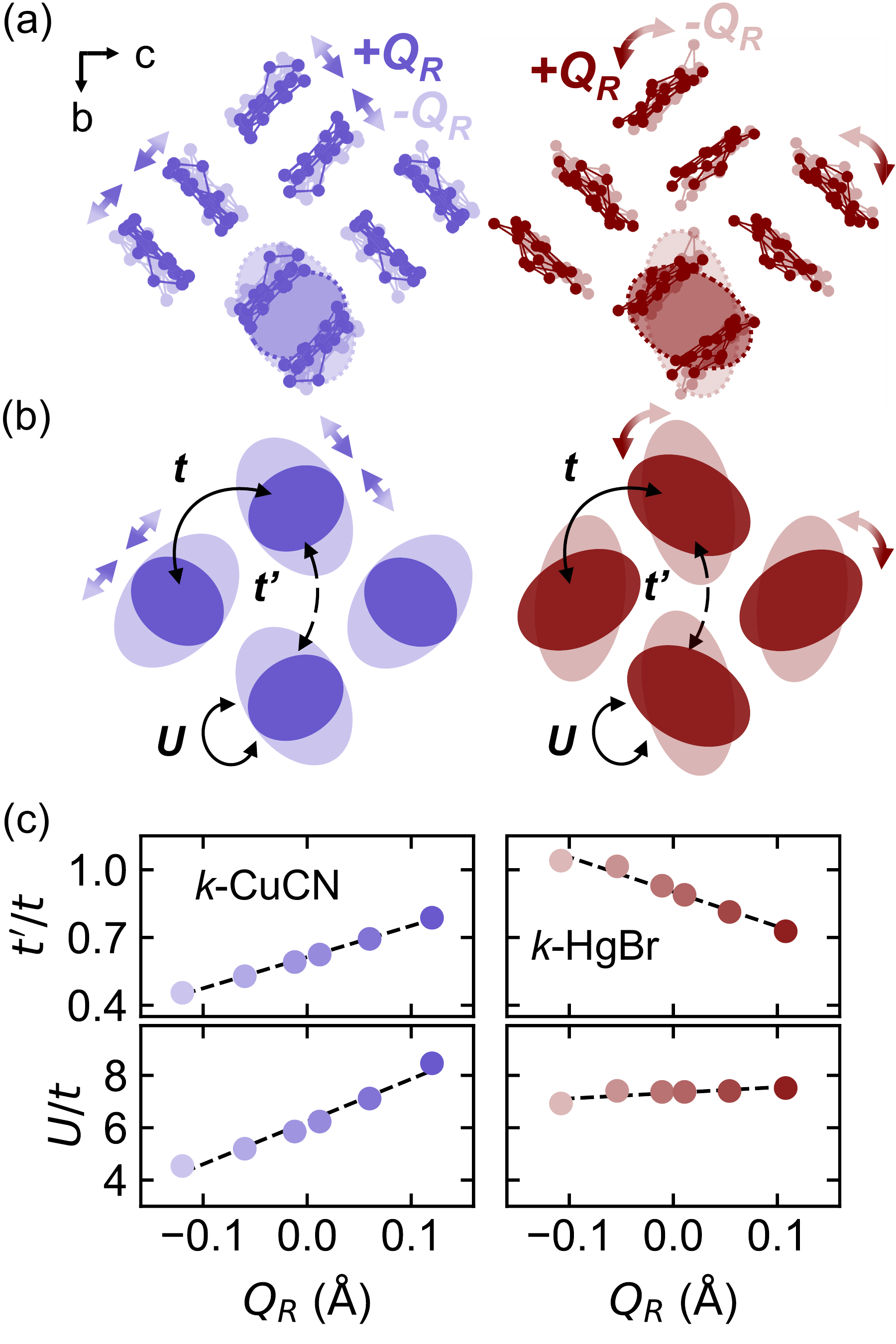}
    \caption{(a) Phonon-displaced (ET)$_2^+$ layers (S and C atoms only) for $\kappa$-CuCN (left) and $\kappa$-HgBr (right). (b) Sketch of the phonon-induced displacements of the (ET)$_2^+$ dimers, featuring intra-dimer distance changes in $\kappa$-CuCN and dimer rotation in $\kappa$-HgBr. The resulting structural distortions modify the Coulomb repulsion $U$ and the inequivalent hopping amplitudes $t$ and $t'$.
    (c) Phonon-induced modulation of geometrical frustration $t'/t$ (top) and effective electronic correlation $U/t$ (bottom) as a function of phonon displacement $Q_R$, obtained from {\it ab initio} band structure calculations \cite{supp}.} 
    \label{fig4}
\end{figure}

This behavior mirrors equilibrium trends associated with anion substitution \cite{Mori2003structural,Mori2006intermolecular,Kandpal2009revision}. Although a direct calculation of the phonon-induced displacement is impractical due to the large unit cell, nonlinear excitation of ET molecular modes for fluence levels comparable to these experiments is known to induce 1–10\% changes in the Hubbard $U$\cite{Singla2015THz,buzzi2020photomolecular}. At equilibrium, such variations in $U$ correspond to intradimer distance changes of approximately 1-10~pm \cite{jeschke2012}. Accordingly, we expect photoinduced modulations of geometric frustration in the 1–10\% range for both compounds.

In summary, we have demonstrated a nonlinear phononic coupling between local molecular vibrations and nonlocal structural modes in $\kappa$-(ET)$_2$X compounds. By resonantly driving ET molecular vibrations associated with the ethylene end groups, we induce a cooperative motion of cation and anion layers, which are expected to modulate correlation strength $U/t$ and geometric frustration ratio $t'/t$. While the dynamical control of electronic interactions has been previously explored \cite{ Kaiser2014optical, Singla2015THz,buzzi2020photomolecular}, the direct modulation of geometric frustration, an essential parameter governing spin dynamics, has so far remained elusive. Our results establish nonlinear phonon excitation as a promising route to dynamically alter magnetic frustration in the quantum spin liquid candidates $\kappa$-CuCN and $\kappa$-HgBr by modulating transfer integrals within their triangular lattices over an extended temperature range. This mechanism could open new control avenues for correlated states in frustrated systems, potentially steering them toward nonequilibrium quantum spin liquid states featuring renormalized spinon and triplon excitations and detectable via mid-infrared optical \cite{Elsasser2012power} and 2D THz spectroscopy \cite{Wan2019resolving}. Moreover, the dynamic tuning of geometric frustration could represent a powerful lever to explore the phase diagram of light-induced superconductivity in $\kappa$-(ET)$_2$X materials \cite{Buzzi2021phase}.

The authors thank N. P. Armitage, P. M. Bonetti, M. Buzzi, N. Drichko, C. Hotta, J. Liebman, M. F\"orst, and H. Padma for insightful discussions. This work was primarily supported by the U.S. Department of Energy, Office of Basic Energy Sciences, Early Career Award Program, under Award No. DE-SC0022883. This study was also supported by the Deutsche Forschungsgemeinschaft (DFG) via DR228-39-3 and DR228-68-1. K. K. acknowledges the support of the JPSJ KAKENHI (21K18144), JSPS Core-to-Core Program (JPJSCCA20240001), and the Alexander von Humboldt Foundation. I. L. acknowledges financial support from the European Regional Development Fund for the "Center of Excellence for Advanced Materials and Sensing Devices" (Grant No. KK.01.1.1.01.0001).

\bibliography{references_arxiv}

\end{document}


\title{Supplementary material:\\ Photoinduced Frustration Modulation in $\kappa$-type Quantum Spin Liquid Candidates}

\author{M. Tepie}
\thanks{These authors contributed equally to this work}
\author{F. Glerean}
\thanks{These authors contributed equally to this work}
\affiliation{Department of Physics, Harvard University, Cambridge, Massachusetts 02138, USA\looseness=-1}

\author{J. Ov\v{c}ar}
\affiliation{Division of Theoretical Physics, Ruđer Bošković Institute, Zagreb 10000, Croatia}

\author{S. Priya}
\affiliation{1. Physikalisches Institut, Universität Stuttgart, Stuttgart 70569, Germany}

\author{K. Miyagawa}
\affiliation{Department of Applied Physics, University of Tokyo, Tokyo 113-8656, Japan}

\author{H. Taniguchi}
\affiliation{Graduate School of Science and Engineering, Saitama University, Saitama 338-8570, Japan}

\author{K. Kanoda}
\affiliation{1. Physikalisches Institut, Universität Stuttgart, Stuttgart 70569, Germany}
\affiliation{Department of Advanced Materials Science, University of Tokyo, Kashiwa 277-8561, Japan}
\affiliation{Max Planck Institute for Solid State Research, 70569 Stuttgart, Germany}

\author{I. Lon\v{c}ari\'{c}}
\affiliation{Division of Theoretical Physics, Ruđer Bošković Institute, Zagreb 10000, Croatia}

\author{M. Dressel}
\affiliation{1. Physikalisches Institut, Universität Stuttgart, Stuttgart 70569, Germany}

\author{M. Mitrano}
\email{mmitrano@fas.harvard.edu}
\affiliation{Department of Physics, Harvard University, Cambridge, Massachusetts 02138, USA\looseness=-1}

\maketitle

\tableofcontents

\section{Sample preparation and equilibrium optical properties}

\begin{figure*}[htbp]
    \centering
    \includegraphics[width=.9\textwidth]{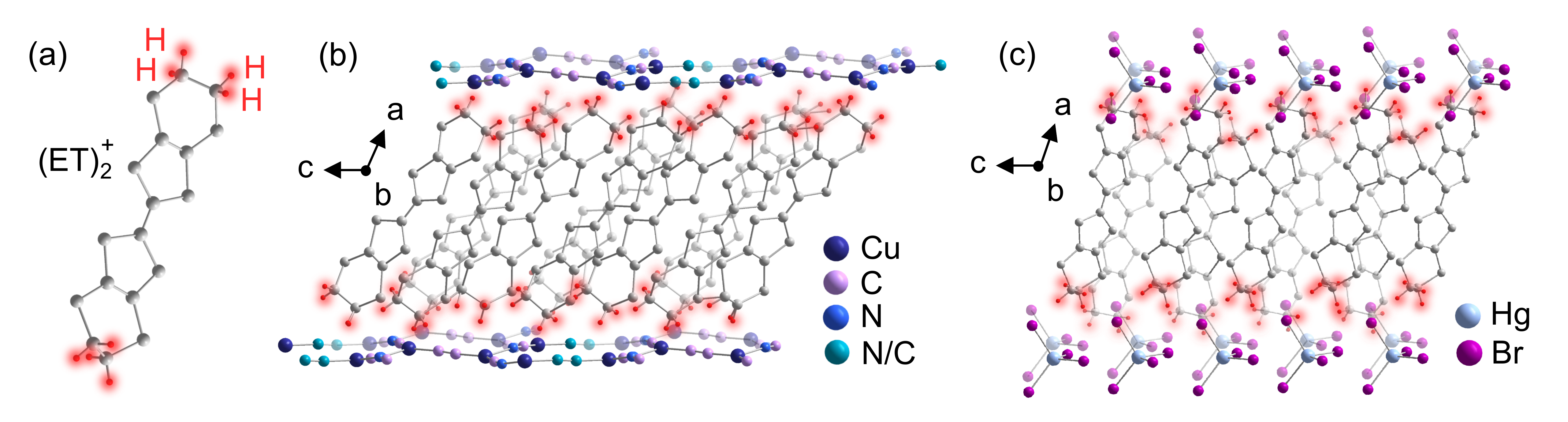}
    \caption{The ethylene end-groups of the BEDT-TTF (ET) molecules (a) couple the ET layers to the anion layers in  $\kappa$-CuCN (b) and $\kappa$-HgBr (c) via hydrogen bonds.}
    \label{fig:crystals}
\end{figure*}

Single crystals of $\kappa$-(BEDT-TTF)$_2$Cu$_2$(CN)$_3$ ($\kappa$-CuCN) and $\kappa$-(BEDT-TTF)$_4$Hg$_{2.89}$Br$_8$ ($\kappa$-HgBr) were grown by electrochemical methods \cite{li1998}. We employed samples with shiny $bc$ surfaces about 1$\times$1 mm$^2$ in size. The typical thickness is 10-50 $\mu$m for $\kappa$-CuCN crystals and 200-300 $\mu$m for $\kappa$-HgBr. For both materials, the $b$ and $c$ axes describe the crystal layers, following the convention in Ref. \cite{li1998}. The samples have a large thermal shrinkage rate and to avoid the formation of cracks we modify the temperature at a rate slower than 1 K/min.

In Fig. \ref{fig:crystals} we show the full crystalline structure of the two compounds. The samples have similar triangular arrangements of molecular dimers separated by anionic layers.
Equilibrium optical properties of the two materials are measured with Fourier-transform infrared (FT-IR) spectroscopy. FT-IR measurements were performed using a Bruker Vertex 80v spectrometer equipped with an Hyperion microscope and cryostat in the range of 600-18000 cm$^{-1}$ (MIR, NIR, VIS) with appropriate combination of source, beamsplitter, optical windows and using a freshly prepared gold mirror (resolution = 2 cm$^{-1}$). A Bruker IFS 113v spectrometer with custom-built cryostat equipped with gold overcoating setup was used to measure reflectivity in the range 100-600 cm$^{-1}$ (FIR range) (resolution= 1 cm$^{-1}$). The reflectivity (> 600 cm$^{-1}$) was normalized by referencing it to FIR range. In order to perform a Kramers-Kronig analysis, the data were extrapolated below and above \cite{tanner2015use} the measured frequency range. Results for $\kappa$-CuCN were reported in \cite{Elsasser2012power}, while we included new original data for the optical properties of $\kappa$-HgBr. 

\section{Pump-probe measurements}

We provide here additional information supporting the pump-probe results presented in Fig. 2 and Fig. 3 of the main text. Unless otherwise stated, we measure the system at 100 K, with 161 meV pump (1 mJ/cm$^2$) polarized // $c$ ($b$) for $\kappa$-CuCN ($\kappa$-HgBr), and 520 meV probe // $b$ for both samples. 

\subsection{Time-dependent reflectivity fits}

\begin{figure}[htbp]
    \centering
    \includegraphics[width=.4\textwidth]{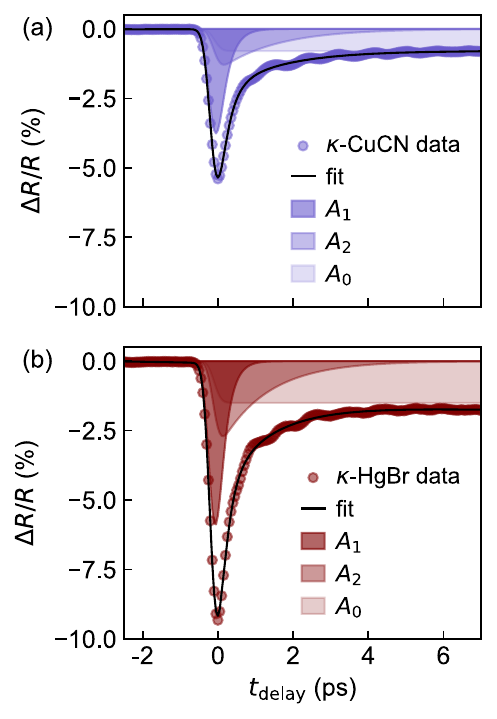}
    \caption{Multi-component fit for $\kappa$-CuCN (a) and $\kappa$-HgBr (b) excited with a 161 meV pump (1 mJ/cm$^2$) at 100 K. The probe is polarized along the $b$-axis, the pump along the $c-$ and $b-$axis, respectively.}
    \label{fig:t_fits}
\end{figure}

We analyze the time dependence of the transient reflectivity $\Delta R(t)$ by performing fits of the exponential decay time using the function 
\begin{equation}\label{eqn:fit_func}
\frac{\Delta R}{R}(t) = \frac{1}{2} \Big[1+\mathrm{erf}\Big(\frac{t-t_0}{\tau_0}\Big)\Big] \Big(A_0 + \sum_{i=1,2} A_ie^{-(\frac{t-t_0}{\tau_i})} \Big),
\end{equation}
where $t_0$ is an arbitrary offset to correct for the pump arrival time and $\tau_0$ is the initial rise time of the response. $\tau_1$ and $\tau_2$ are two characteristic decay timescales, while $A_0$ accounts for longer relaxation processes. Figure \ref{fig:t_fits} shows the fits of the curves presented in Fig.~2 of the main text, and the fit parameters are reported in Table \ref{tab:fits}.

\begin{table}[htbp]
\centering
\caption{Fit parameters for Fig. \ref{fig:t_fits}.}
\label{tab:fits}
\begin{tabular}{p{1.5cm}p{3cm}p{3cm}}
                    \toprule
                    & \multicolumn{1}{l}{$\kappa$-CuCN} & \multicolumn{1}{l}{$\kappa$-HgBr}  
 \\ \midrule
$\tau_0$ (ps)           & 0.216 $\pm$ 0.002& 0.205 $\pm$ 0.002\\
$\tau_1$ (ps)           & 0.246 $\pm$ 0.010& 0.219 $\pm$ 0.013\\
$\tau_2$ (ps)           & 1.500 $\pm$ 0.118& 1.265 $\pm$ 0.087\\
$A_0$            & -0.0080 $\pm$ 0.0006& -0.0150 $\pm$ 0.0008\\
$A_1$           & -0.0733 $\pm$ 0.0009& -0.1159 $\pm$ 0.0021\\
$A_2$            & -0.0161 $\pm$ 0.0007& -0.0344 $\pm$ 0.0015\\
\bottomrule
\end{tabular}
\end{table}

\subsection{Photosusceptibility analysis}

To analyze the energy dependence of the transient optical response of the dimers, we correct the transient reflectivity amplitude $\Delta R/R$ to account for the number of pump photons effectively absorbed by the samples. 
\begin{figure}[H]
\includegraphics[width=.5\textwidth]{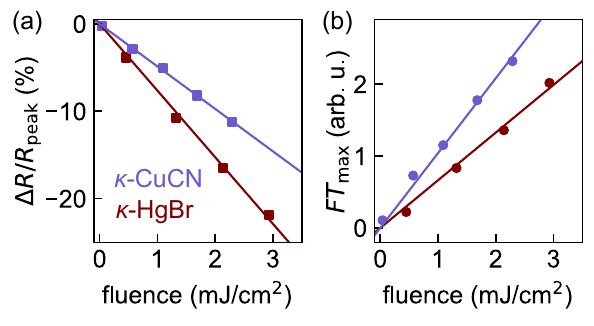}
  \caption{Linear pump fluence dependence of the peak reflectivity change (a) and of the Fourier-transform amplitude maximum (b), for $\kappa$-CuCN and $\kappa$-HgBr.\label{fig:fluence}}
\end{figure}
Since the pump-induced $\Delta R/R$ scales linearly with fluence (see Fig. \ref{fig:fluence}), we define an effective absorbed fluence $f=F(1-R)$ where the input fluence $F$ is rescaled accounting for reflective losses \cite{Caviglia2012ultrafast}. Figures \ref{fig:peak_wlpol_dep}(a),(b) present the results of this analysis for the peak transient reflectivity response, demonstrating that its exponential decay follows the absorption profile in the energy range of the ethylene end group vibrations. 
\begin{figure}[htbp]
\includegraphics[width=.49\textwidth]{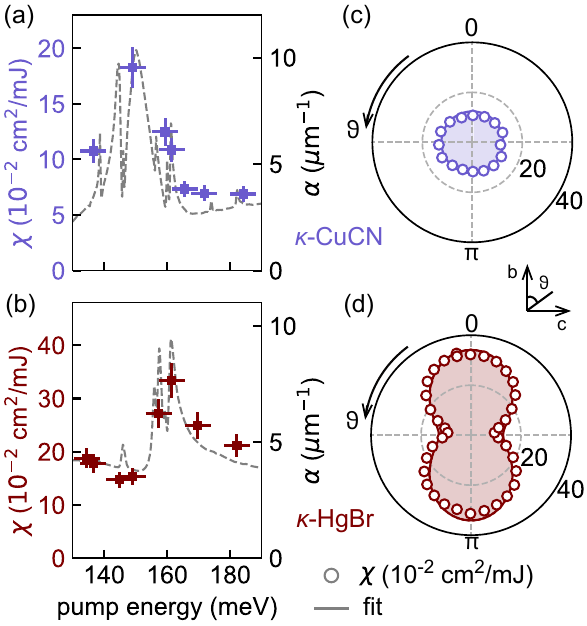}
  \caption{Effective photosusceptibility as a function of pump photon energy for $\kappa$-CuCN (a) and $\kappa$-HgBr (b). The excited reflectivity response (symbols) resonates with the ethylene molecular modes observed in linear absorption (dashed lines). The pump is polarized along the $c$- and $b$-axes, respectively. The probe polarization dependence of the photosusceptibility (measured with 161 meV pump along $b$) is isotropic for $\kappa$-CuCN (c) and anisotropic for $\kappa$-HgBr (d), reflecting the distinct anion lattice symmetries.\label{fig:peak_wlpol_dep}}
\end{figure}
In the main text, we analyze the phonon response by calculating $\chi_{ph}$ from the amplitude of the Fourier transform (FT) of the coherent oscillations corrected for $f$. To preserve the physical dimensions of $\chi_{\mathrm{ph}}$, we use  $\chi_{\mathrm{ph}}$  = $(FT[(\Delta R/R)_{ph}]_{\mathrm{max}}\times\frac{\mathrm{t_{step}} }{(\mathrm{t_{range}/2)}})/f$, where $\mathrm{t_{step}}$ is the step size of the oscillation time axis and $\mathrm{t_{range}}$ is the extension of the time window used for the FT.
To analyze the symmetries of  $\chi$ and $\chi_{ph}$, we fit the probe polarization dependence with\begin{equation}\label{eqn:polar fit_func}
 A_{C_2} \sin^2{(\vartheta+\phi)} + A_{C_4} \sin^2{(2\vartheta+\phi)} +A_{C_{\infty}} 
\end{equation}
to describe $C_2$, $C_4$ and isotropic components. The response, in Fig. \ref{fig:peak_wlpol_dep}(c),(d) and in the main text, is always isotropic for $\kappa$-CuCN, while it presents $C_2$ and $C_4$ symmetry components in $\kappa$-HgBr.

\subsection{Temperature dependent measurements}
\begin{figure}[htbp]
    \centering
    \includegraphics[width=.49\textwidth]{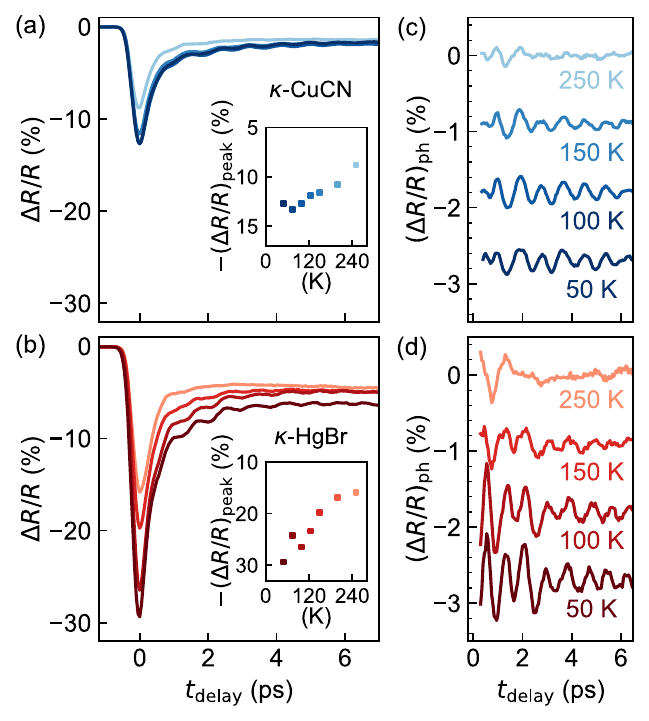}
    \caption{Temperature-dependent transient reflectivity $\Delta R/R$ for $\kappa$-CuCN (a) and $\kappa$-HgBr (b), employing pump fluence of 2.3 mJ/cm$^2$ and 2.8 mJ/cm$^2$ respectively. Insets show the peak response as a function of temperature. (c),(d) Temperature-dependent phonon oscillations after subtraction of exponential electronic background. Curves are vertically shifted for clarity. }
    \label{fig:temperature}
\end{figure}

The transient reflectivity change in both $\kappa$-CuCN and $\kappa$-HgBr gradually increases as temperature is lowered [Fig. \ref{fig:temperature}]. The $\sim$1.2-THz phonon oscillations are coherent at lower temperatures. As temperature increases, the phonon lifetime decreases and the oscillations almost completely disappear above 250 K.

\section{Phonon-modulated electronic band structures }

\begin{figure}[htbp]
    \centering
    \includegraphics[width=.4\textwidth]{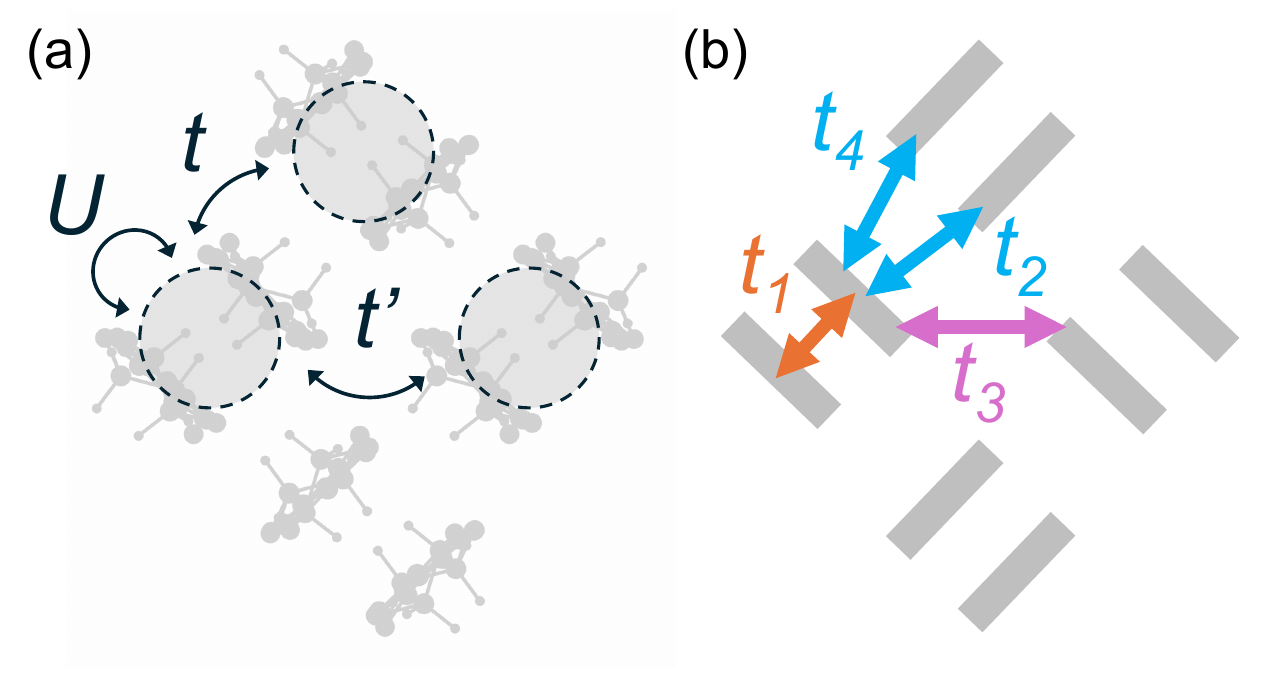}
    \caption{(a) Hubbard parameters describing the on-dimer repulsion $U$ and inter-dimer hoppings $t, t'$ on the the $\kappa$-(ET)$_2$ layer. (b) Schematic representation of the intermolecular transfer parameters employed to model the electronic interactions.}
    \label{fig:model}
\end{figure}

The low-energy electronic band structure of the $\kappa$-(BEDT-TTF)$_2$X charge-transfer salts is well captured by the four-band, 3/4-filled Hubbard Hamiltonian \cite{Kandpal2009revision, koretsune2014evaluating, guterding2016}:

\begin{equation}\label{eqn:hubbard_hamiltonian}
    H = \sum_{ij\sigma}t_{ij}\big(c^\dag_{i\sigma}c_{j\sigma} + \text{h.c.}\big) + \frac{U}{2}\sum_{i\sigma}n_{i\sigma}n_{i\bar{\sigma}},
\end{equation}

where $c^\dag_{i\sigma}$ ($c_{i\sigma}$) create (annihilate) an electron with spin $\sigma$ in an orbital of the $i^{\mathrm{th}}$ BEDT-TTF molecule, $t_{ij}$ are the intermolecular hopping parameters, $U$ is the Hubbard parameter and $n_{i\sigma}$ = $c^\dag_{i\sigma}c_{i\sigma}$ (see Fig. \ref{fig:model}).

By considering the four largest hopping parameters $(t_1, t_2, t_3, t_4)$, we adopt an effective model in which hopping between the (BEDT-TTF)$_2$ dimers is defined by the following geometric formulas \cite{tamura1991}:

\begin{eqnarray}\label{eqn:geometric_formulas}
    t &= \frac{|t_2| + |t_4|}{2}, \\
    t' &= \frac{|t_3|}{2}.
\end{eqnarray}
The Hubbard $U$ parameter is given by $U\approx 2|t_1|$. 

In this work, we aim to estimate the variation of $t_{ij}$, $t$, $t'$ and $U$ in $\kappa$-CuCN and $\kappa$-HgBr as a consequence of phonon excitations. To this end, we employ density-functional theory (DFT) calculations in the frozen-phonon approximation. In Section \ref{sec:computational_details}, we describe the computational methods, while we discuss the results for $\kappa$-CuCN and $\kappa$-HgBr in Sections \ref{sec:kappa_cn} and \ref{sec:kappa_hg}, respectively.

\subsection{Computational method}\label{sec:computational_details}

To perform DFT calculations, we employed the open-source Quantum ESPRESSO suite for quantum simulations of materials \cite{giannozzi2009, giannozzi2017}. We used the Garrity-Bennett-Rabe-Vanderbilt (GBRV) pseudopotentials \cite{garrity2014} and the vdW-DF-cx \cite{thonhauser2007, langreth2009, thonhauser2015, berland2015} exchange-correlation functional, with the kinetic energy cutoff of the plane-wave basis set equal to 90 Ry. The self-consistent field calculations were performed on a $1\times 3\times 2$ (for $\kappa$-CuCN) or $3\times 2\times 1$ (for $\kappa$-HgBr) Monkhorst-Pack $k$-point grids \cite{monkhorst1976}, while the non-self-consistent field calculations were performed on uniform $6\times 6\times 6$ grids. All atomic positions and cell parameters were relaxed until the force on each ion was less than $10^{-4}$ Ry/Bohr and the potential energy changed less than $10^{-5}$ eV/atom between subsequent optimization steps. The target pressure was set to 0 kbar with a convergence threshold of 0.5 kbar.

We obtained the phonon modes at the $\Gamma$ point in the harmonic approximation using the finite-differences method as implemented in the open-source phonopy software package \cite{togo2023implementation, togo2023first}. We then modulated the relaxed atomic structure along two directions of the selected phonon modes by displacing the $i^{\mathrm{th}}$ atom in the structure using the formula:

\begin{equation}\label{eqn:phonon_displacement}
    \Vec{Q}_i = \pm \frac{\Tilde{Q}}{\sqrt{N m_i}}\Vec{e}_i,
\end{equation}

where $N$ is the total number of atoms, $m_i$ is the mass of the $i^{\mathrm{th}}$ atom and $\Vec{e}_i$ is the $i^{\mathrm{th}}$ part of the mode eigenvector. The $+$ and $-$ signs correspond to the two investigated directions of a selected phonon mode. We calculated the amplitude $\Tilde{Q}$ as:

\begin{equation}\label{eqn:modulation_amplitude}
    \Tilde{Q} = \frac{(N-N_\text{H})\sqrt{N}}{\sum_{i\neq \text{H}}|\Vec{e}_i| /\sqrt{m_i}} n Q_0,
\end{equation}

where $N_\text{H}$ is the number of hydrogen atoms and the sum in the denominator runs over all atoms except hydrogens. By using equation \eqref{eqn:modulation_amplitude}, we ensure that the average displacement $Q$ of the atoms in the modulated structures, disregarding hydrogen atoms, is equal to $nQ_0$. We create modulated structures for typical displacements on the order of picometers (as observed in non-equilibrium \cite{Singla2015THz,buzzi2020photomolecular} and temperature dependent studies \cite{jeschke2012}) using $n=(1, 5, 10)$ and $Q_0=1.20$ pm for $\kappa$-CuCN and $Q_0=1.08$ pm for $\kappa$-HgBr. 

For each equilibrium (relaxed) and modulated structure, we constructed four Wannier functions centered on the BEDT-TTF molecules using the open-source Wannier90 package \cite{mostofi2014}. By including the two bands crossing the Fermi level, for $\kappa$-CuCN we performed a disentanglement procedure on the top 6 valence bands, while for $\kappa$-HgBr we considered only the top 4 valence bands, as they are well isolated from the lower-energy valence bands. We obtained excellent fits for the low-energy DFT band structures by specifying the initial projection functions for the unitary transformation matrix as $p_z$ orbitals centered at BEDT-TTF centers of mass, with $z$ oriented perpendicularly to the plane fitted to the positions of the atoms in the BEDT-TTF molecules. The hopping parameters $(t_1, t_2, t_3, t_4)$ are equal to the matrix elements of the Hamiltonian written in the basis of the four Wannier functions.

\subsection{$\kappa$-(BEDT-TTF)$_2$Cu$_2$(CN)$_3$}\label{sec:kappa_cn}

The hopping parameters $(t_1, t_2, t_3, t_4)$, $t$ and $t'$ for $\kappa$-CuCN have been calculated using \textit{ab initio} methods in several works \cite{Kandpal2009revision, nakamura2009, jeschke2010, jeschke2012, koretsune2014evaluating, guterding2015, guterding2016}. As shown by Koretsune \textit{et al.} \cite{koretsune2014evaluating}, a seemingly small change (about 2\%) in atomic positions and cell parameters, resulting from \textit{ab initio} optimization of the experimental crystal structure, may lead to a large variation (more than 25\%) of the hopping integrals. For this reason, in the majority of prior studies, the hopping integrals were evaluated either using the unrelaxed or partially relaxed crystal structures. In the following paragraph, we review results from previous studies and for the structures used therein.

\begin{figure*}[htbp]
    \centering
    \includegraphics[width=0.9\linewidth]{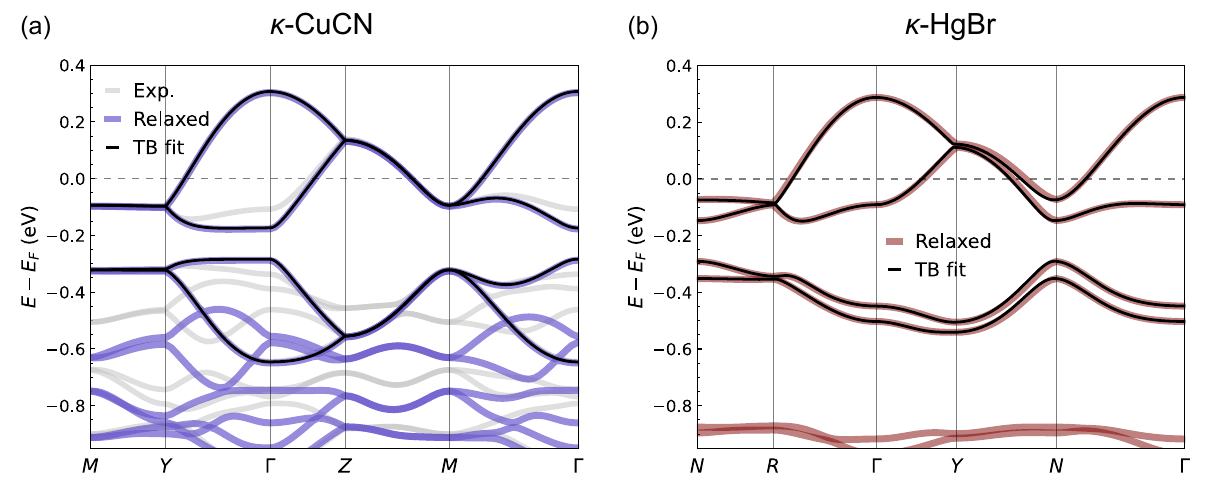}
    \caption{Equilibrium electronic bands calculated for the relaxed crystal structures and their tight binding fits of the low energy bands, for $\kappa$-CuCN (a) and $\kappa$-HgBr (b). For $\kappa$-CuCN, we include the electronic bands obtained from the $T=20$ K experimental structure reported in Ref. \cite{jeschke2012}.}
    \label{fig:eq_bands}
\end{figure*}

Kandpal \textit{et al.} \cite{Kandpal2009revision} have based calculations on a crystal structure obtained experimentally at ambient pressure \cite{geiser1991}, with the hydrogen positions computationally found using the Car-Parrinello projector-augmented wave molecular dynamics (CPMD). To model structures at low and high pressures, the experimental structure was then either partially relaxed, with the positions of S and Cu atoms and the lattice parameters kept fixed, or fully relaxed, with the pressure being constrained to 0.75 GPa. Nakamura \textit{et al.} \cite{nakamura2009} used the same experimental crystal structure \cite{geiser1991}. Jeschke \textit{et al.} \cite{jeschke2010} also found the hydrogen positions, which were lacking in the experimentally solved structure \cite{geiser1991}, using CPMD and compared the hopping integrals for several computational setups. In a combined experimental and theoretical study by Schlueter \textit{et al.} \cite{jeschke2012}, the crystal structure of $\kappa$-CuCN was solved for a series of temperatures between 5 K and 300 K and the temperature dependence of the hopping integrals was calculated for these experimental structures. Koretsune \textit{et al.} \cite{koretsune2014evaluating} performed a detailed comparative study, elucidating the dependence of the hopping parameters on different crystal structures \cite{geiser1991, jeschke2012}, crystal symmetries and geometry relaxation. Guterding \textit{et al.} \cite{guterding2015} calculated the hopping integrals for various conformations of the BEDT-TTF molecules by relaxing the ethylene end groups and the two neighboring S atoms.

As we are interested in electronic structure changes induced by phonon excitation, we must first relax the crystal structure to a local minimum to be able to calculate the interatomic force constants. For the initial structure, we use the crystal structure obtained by Schlueter \textit{et al.} \cite{jeschke2012} at $T=20$ K \cite{ccdc2012right}, as the Cambridge Structural Database (CSD) \cite{groom2016} entry of the structure obtained at $T=5$ K, at the time of writing this work, does not correspond to the structure described in the original publication \cite{ccdc2012wrong, jeschke2012}. The symmetry of the experimental structure is $P2_1/c$ \cite{jeschke2012}. We select the orientation of the CN$^-$ group in the inversion center corresponding to $P_c$ symmetry.

A comparison of the initial and relaxed unit cell parameters in this work and Koretsune \textit{et al.} \cite{koretsune2014evaluating} is given in Table \ref{tab:cell_opt}. As we discuss below, the comparison with Ref. \cite{koretsune2014evaluating} is relevant since it is the only investigation in which the hopping parameters were calculated for a fully relaxed structure at low pressure. We find that the cell parameters we obtained after optimization are in good agreement with the initial (experimental) parameters for $T=20$ K. We may ascribe this agreement to the inclusion of the Van der Waals (VdW) interactions in the calculations by using the vdw-df-cx exchange correlation functional \cite{thonhauser2007, langreth2009, thonhauser2015, berland2015} (for more details, see Sec. \ref{sec:computational_details}), since VdW interactions have been shown to be pivotal to the accuracy of modeling the physics of $\kappa$-(BEDT-TTF)$_2$X salts \cite{lazic2019}. The optimization procedure in Koretsune \textit{et al.} \cite{koretsune2014evaluating} led to an expansion of the experimental unit cell; most likely due to the use of the Perdew-Burke-Ernzerhof (PBE) \cite{perdew1996} exchange-correlation functional.

\begin{table*}[t]
\centering
\caption{Unit cell parameters for the $\kappa$-CuCN crystal structures. The $\kappa$-CuCN initial experimental structures at different temperatures are obtained from Ref. \cite{jeschke2012}. As the initial structure, we used the $T=20$ K experimental structure. We include for comparison the relaxed structure obtained by Koretsune \textit{et al.} \cite{koretsune2014evaluating}, starting from the $T=5$ K structure. Note that the cell angles $\alpha=\gamma=90$\textdegree.}
\label{tab:cell_opt}
\begin{tabular}{p{2cm}p{2cm}p{2cm}p{2cm}p{2cm}}
                    \toprule
                    & \multicolumn{1}{l}{Exp. (20 K)} & \multicolumn{1}{l}{Relaxed} & \multicolumn{1}{l}{Exp. (5 K)} & \multicolumn{1}{l}{Relaxed}\\
                    & \multicolumn{1}{l}{(Ref. \cite{jeschke2012})} & \multicolumn{1}{l}{(this work)} & \multicolumn{1}{l}{(Ref. \cite{jeschke2012})} & \multicolumn{1}{l}{(Ref. \cite{koretsune2014evaluating})}\\ 
                    \midrule
$a$ (\AA)           & 16.072    & 15.979   & 16.062  & 16.45   \\
$b$ (\AA)           & 8.536     & 8.580    & 8.544   & 8.83      \\
$c$ (\AA)           & 13.262    & 13.061   & 13.271  & 13.76       \\
$\beta$ (\textdegree) & 115.088 & 115.056  & 115.093 & 112.47 \\
\bottomrule
\end{tabular}
\end{table*}
\begin{table*}[t]
\centering
\caption{Hopping parameters calculated for $\kappa$-CuCN and $\kappa$-HgBr. For $\kappa$-CuCN, we include for comparison the parameters obtained for the initial and relaxed structures in this work and  by Koretsune \textit{et al.} \cite{koretsune2014evaluating}. }
\label{tab:hopping_parameters}
\begin{tabular}{p{2.5cm}p{2cm}p{2cm}p{2cm}p{3cm}p{2cm}}
                    \toprule
                    & \multicolumn{4}{l}{$\kappa$-CuCN}  & \multicolumn{1}{l}{$\kappa$-HgBr} \\ 
                    \midrule
                    & \multicolumn{1}{l}{Init. (20 K)} & \multicolumn{1}{l}{Relaxed} & \multicolumn{1}{l}{Init. (5 K)} & \multicolumn{1}{l}{Relaxed} & \multicolumn{1}{l}{Relaxed}  \\ 
                    & \multicolumn{1}{l}{(This work)} & \multicolumn{1}{l}{(This work)} & \multicolumn{1}{l}{(Ref. \cite{koretsune2014evaluating})} & \multicolumn{1}{l}{(Ref. \cite{koretsune2014evaluating})}&  \multicolumn{1}{l}{(This work)}\\ 
                    \midrule
$t_1$ (meV)          & 204    & 188   & 199  & 146   & 201    \\
$t_2$ (meV)          & 87     & 107    & 85   & 65   & 63    \\
$t_3$ (meV)          & 93    & 76   & 91  & 65   & 99     \\
$t_4$ (meV)          & 18    & 18   & 17  & 30   & 45     \\
$t'/t$          & 0.89    & 0.61   & 0.89  & 0.68  & 0.91\\
\bottomrule
\end{tabular}
\end{table*}

A comparison of the calculated electronic band structures for the initial and relaxed structures is shown on Fig. \ref{fig:eq_bands}(a). The band structure around the Fermi level changes significantly as a consequence of the seemingly small variation of the geometric parameters between the initial and relaxed structures. As discussed below, and as was previously observed in the literature \cite{koretsune2014evaluating}, this variation in the band structure also leads to a large deviation in the hopping parameters.
In Fig. \ref{fig:eq_bands}(a), we show the tight-binding (TB) fit (corresponding to the kinetic part of the Hamiltonian (\ref{eqn:hubbard_hamiltonian})) of the low-energy band structure obtained by constructing Wannier functions, as described in Sec. \ref{sec:computational_details}. The agreement between the band structure and the TB fit is excellent. To validate our fitting method and estimate the effects of crystal structure optimization, we performed the same fitting procedure on the experimental structure obtained at $T=20$ K. The results for the relevant hopping parameters are shown in Table \ref{tab:hopping_parameters}, along with a comparison to the values obtained by Koretsune \textit{et al.} \cite{koretsune2014evaluating}. The values of the hopping parameters obtained for the respective experimental structures are in close agreement, with the small discrepancies being attributable to the differences in the structures at 5 and 20~K and exchange-correlation functionals used in this work and Ref. \cite{koretsune2014evaluating}. On the other hand, there are significant discrepancies in the calculated parameters for the relaxed structures, consistent with the differences in the unit cell parameters given in Table \ref{tab:cell_opt} and the aforementioned sensitivity of the hopping integrals to the crystal structure.

\begin{figure*}[htbp]
    \centering
    \includegraphics[width=1.0\linewidth]{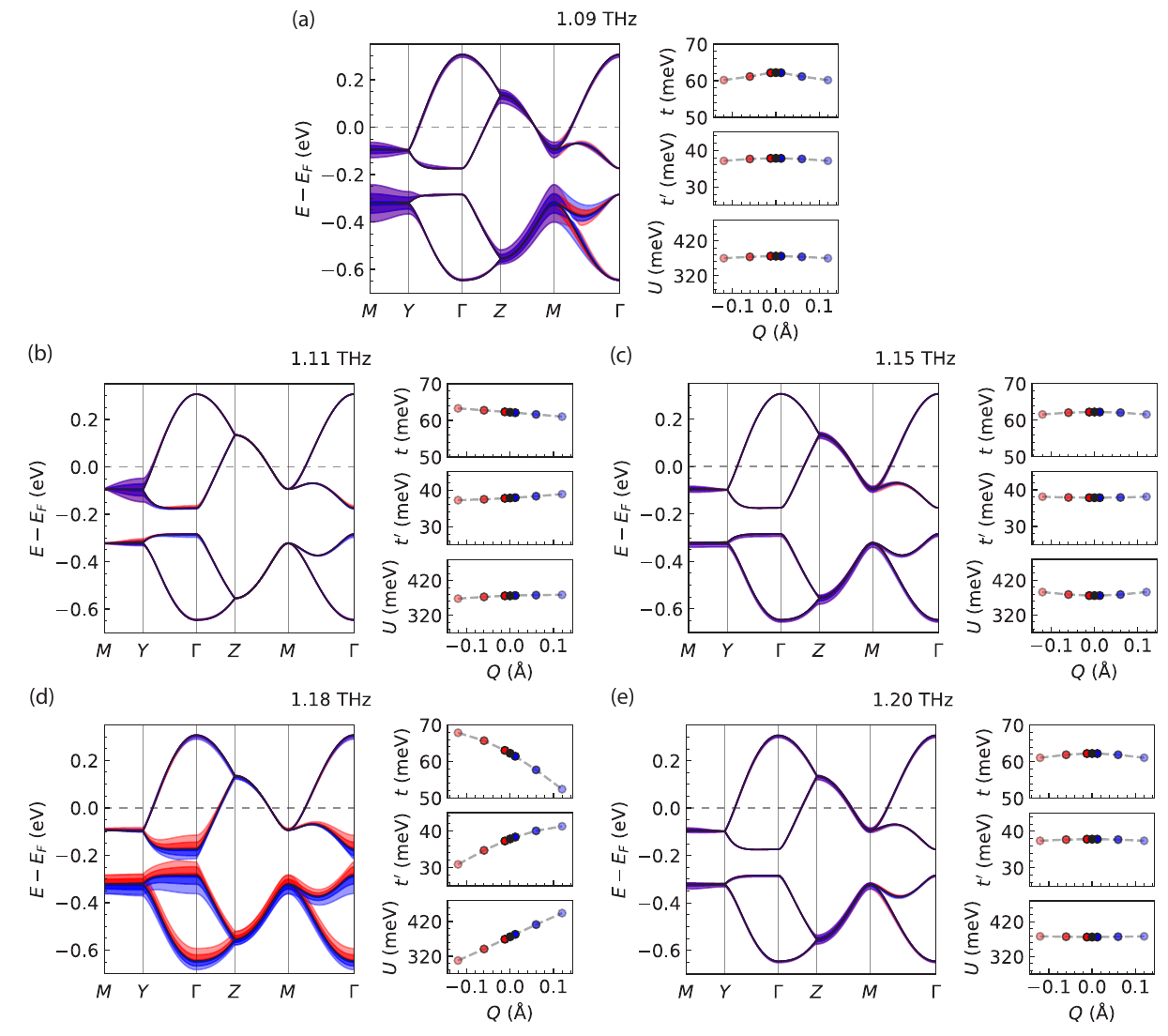}
    \caption{Modulation of electronic properties of $\kappa$-(BEDT-TTF)$_2$Cu$_2$(CN)$_3$ for displacements along the calculated phonon modes with frequencies near 1.12 THz: (a) 1.09 THz, (b) 1.11 THz, (c) 1.15 THz, (d) 1.18 THz, (e) 1.20 THz.
    TB electronic band structures at equilibrium and for structures modulated with different phonon amplitudes (left panel); the dependence of the ratio of the hopping integrals $t$, $t'$ and the Hubbard $U$ parameter on the average modulation amplitude $Q$ (right panels). Shaded areas in the band structures correspond to differences between the modulated and equilibrium bands.}
    \label{fig:kCN_phonon_bands}
\end{figure*}

We now discuss the effect of the light-induced phonon displacement on the hopping parameters. We calculated the hopping parameters for modulations along five phonon modes with frequencies near 1.12 THz (for computational details, see Sec. \ref{sec:computational_details}). The TB band structures for the equilibrium and modulated structures are shown in Fig. \ref{fig:kCN_phonon_bands}(a-e), together with the corresponding dimer hopping integrals $t$, $t'$ and the Hubbard repulsion $U$. 

Among the displacements in this energy range, by far the largest oscillation of the TB parameters is obtained for the mode with a frequency of 1.18 THz [Fig. \ref{fig:kCN_phonon_bands}(d)], which dominates the nonequilibrium response.

\subsection{$\kappa$-(BEDT-TTF)$_4$Hg$_{2.89}$Br$_8$}\label{sec:kappa_hg}

\begin{figure*}[h]
    \centering
    \includegraphics[width=1.0\linewidth]{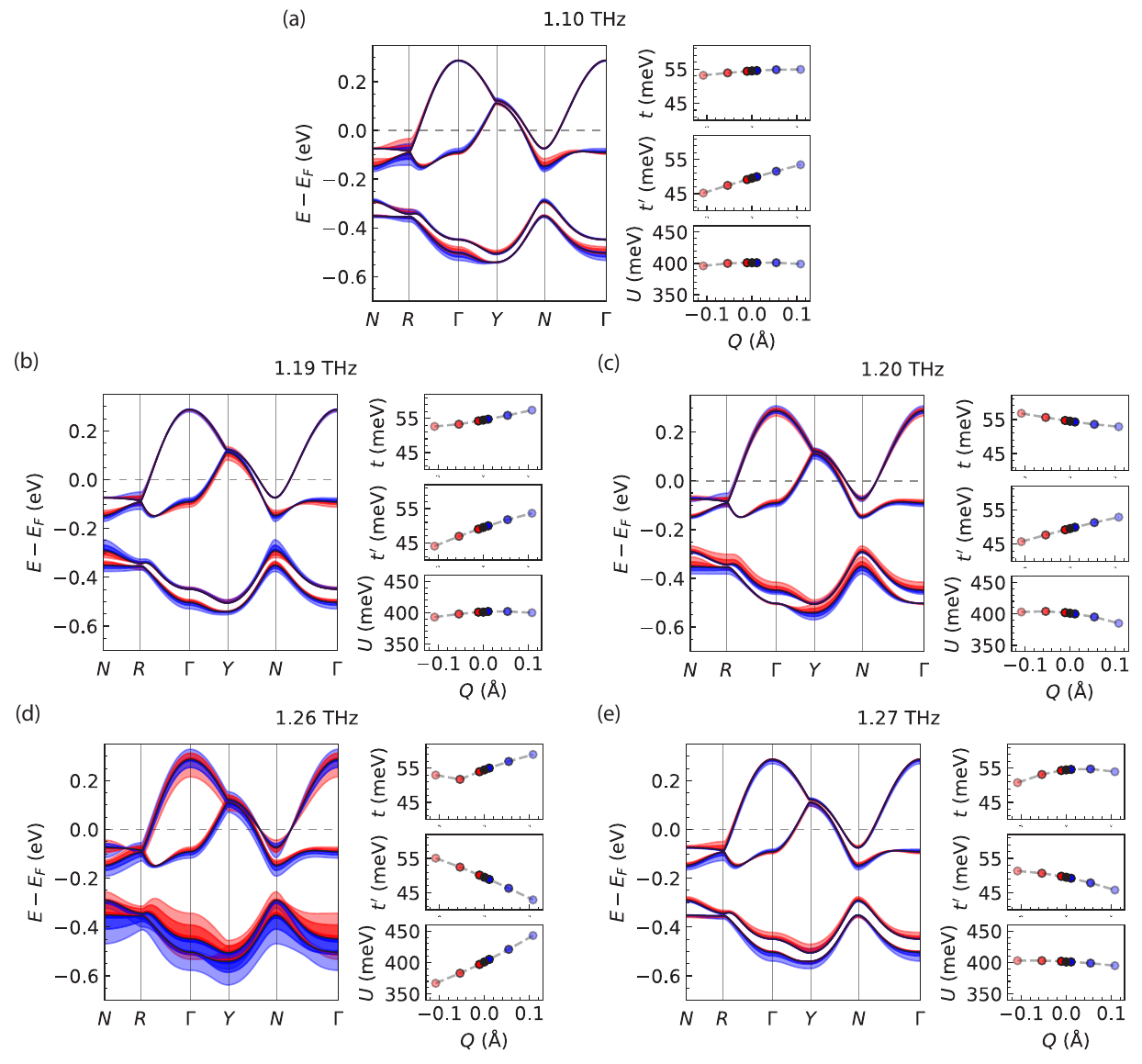}
    \caption{Modulation of electronic properties of $\kappa$-(BEDT-TTF)$_4$Hg$_{2.89}$Br$_8$ for displacements along the calculated phonon modes with frequencies near 1.1 THz: (a) 1.10 THz, (b) 1.19 THz, (c) 1.20 THz, (d) 1.26 THz, (e) 1.27 THz.
    TB electronic band structures at equilibrium and for structures modulated with different phonon amplitudes (left panel); the dependence of the ratio of the hopping integrals $t$, $t'$ and the Hubbard $U$ parameter on the average modulation amplitude $Q$ (right panels). Shaded areas in the band structures correspond to differences between the modulated and equilibrium bands.}
    \label{fig:kHg_phonon_bands}
\end{figure*}
As opposed to $\kappa$-CuCN, for which several computational investigations have been performed in previous studies, no previous \textit{ab initio} calculations on $\kappa$-HgBr can be found in the literature. This is due to the incommensurate two-sublattice crystal structure of the material \cite{li1998}, making precise DFT calculations computationally unfeasible. In this work, we modeled the $\kappa$-(BEDT-TTF)$_4$Hg$_{2.89}$Br$_8$ crystal structure starting from the crystal structure obtained at room temperature \cite{li1998} and constructing a primitive cell from the $\kappa$-(BEDT-TTF)$_2$Br$_8$ sublattice. Subsequently, we added three Hg atoms into the primitive cell at positions we approximated by inspecting the original Hg$_2$ sublattice, finally resulting in a crystal structure containing 115 atoms. The initial and relaxed model $\kappa$-HgBr structures are given in CIF format in the Supplemental Data. The relaxed band structure is shown in Fig. \ref{fig:eq_bands}(b) together with the relative TB fit. The hopping parameters derived from the fit are included in Table \ref{tab:hopping_parameters}.

To calculate the dependence of hopping parameters of the phonon-displaced structure, we then followed the same procedure used for $\kappa$-CuCN and described in Sec. \ref{sec:computational_details}. The corresponding $\kappa$-HgBr phonon-dependent Hubbard parameters are shown in Fig. \ref{fig:kHg_phonon_bands}(a-e). Compared to $\kappa$-CuCN, we observe a distinctly different dispersion of the valence bands and distinguishable oscillation of the hopping parameters for all investigated phonon excitations. Nevertheless, the 1.26 THz mode [Fig. \ref{fig:kHg_phonon_bands}(d)] shows the largest effect on the $U$ and the ratio $t'/t$ and dominates the nonequilibrium response. We note that, due to the aforementioned sensitivity of the hopping integrals to the details of the crystal structure \cite{koretsune2014evaluating}, a more detailed investigation is warranted to validate the accuracy of the employed model of the $\kappa$-HgBr crystal structure.

\clearpage

\bibliography{references_arxiv}